\documentclass[]{article}
\begin{document}
\begin{center}
{\bf \large Operator Approach to Isospin Violation in Pion Photoproduction} 
\vskip 1cm
B. Ananthanarayan\\
Centre for High Energy Physics\\
Indian Institute of Science\\
Bangalore 560 012, India\\
\bigskip

\end{center}
\begin{abstract}
Unambiguous isospin violation in the strong interaction
sector is a key issue in low energy hadronic physics, both
experimentally and theoretically.  Bernstein has
employed the Fermi-Watson theorem to demonstrate that
pion photoproduction is a process where isospin violation 
in the $\pi N$ system can be revealed,  an approach we review here.
Here we propose a general operator approach to the phenomenon in 
pion photoproduction, thereby providing an 
analogue for the framework that was proposed for $\pi N$ scattering
by Kaufmann and Gibbs.  The resulting set of amplitudes could
form the basis for determining the multipole amplitudes
for photoproduction. Thus, the so resulting
phase shift determination from photoproduction can then
be used via the Fermi-Watson theorem to resolve discrepancies
in $\pi N$ phase shift analyses.  We point out that casting effective
Lagrangian results in terms of our framework would be beneficial.  
The upcoming polarization
experiments are an ideal setting to test our approach, 
and also to constrain better the isotensor currents
which strictly are not forbidden.  
\end{abstract}


\bigskip


\section{Introduction}
Of the key issues in hadronic physics (for a eponymous
white paper on the subject, see ref.~\cite{Key}),
the issue of isospin violation in the low energy strong
interaction sector is an important one
(see Sec. 5.1 of~\cite{Key}), both theoretically and
experimentally.    
Charge symmetry breaking in hadronic reactions (for a review
see, ref.~\cite{MNS,Gardestig2}) for which there have been remarkable
new experimental signatures in the reactions 
$d\, d\to\alpha\, \pi^0$~\cite{PRLTA.91.142302} and in the reaction
$n\, p\to d\, \pi^0$~\cite{PRLTA.91.212302}.  Related investigations
in the theoretical front have also been presented, see, ref.~\cite{Gardestig}.  
In general the issue of charge symmetry breaking and isospin conservation
(also known as charge independence in the hadronic sector) and violation
have consequences in the nucleon-nucleon sector,
for a sample of reviews see, e.g., ref.~\cite{Nuclear1,Nuclear2,Ericson}.

The situation is less clear in scattering processes:
there have been several analyses of $\pi N$ scattering data
with the aim of establishing isospin violation in the 
system~\cite{Gibbs_Li_Kaufmann,Matsinos}, which all see
evidence for isospin violation, although the numerical
size of the violation remains uncertain.  This matter is
of great importance to low energy strong interaction
dynamics as pointed out by Weinberg~\cite{Weinberg1}. 
Indeed, isospin violation due to the quark mass difference $m_d-m_u$,
where $m_d$ and $m_u$ are fundamental parameters of
the standard model~\cite{GL_report}, will lead 
to pronounced effects in the $\pi N$ scattering lengths at leading order
in chiral perturbation theory, recalling here that
chiral perturbation theory is the effective low energy theory
of the strong interactions (for a few excellent reviews, see, 
e.g.~\cite{CHPTReviews}).  Mei{\ss}ner and co-workers have 
worked out the consequences of the quark mass difference to higher
order in a series of investigations~\cite{Meissneretal}.

A completely general approach to isospin violation $\pi N$ scattering has
been presented in ref.~\cite{KG}.  In this framework,
which is based on the treatment of all possible operators that
may arise due to isospin violation are considered and classified
thoroughly in terms of operators, denoted by $\theta_i,\, i=1,...10$. 
It thus provides a general platform for the analysis of
the $\pi N$ system.  

In QCD isospin violation is due to the non-vanishing
of $(m_d-m_u)$, and is introduced at the level of the microscopic
Lagrangian as an isovector.  Isospin violation would arise in all 
hadronic processes involving the strong interactions, 
as well as the electromagnetic interactions.  
This could, in principle, at higher orders generate
all possible isospin violating operators in each process of interest, 
that could connect the particles in the initial state to those in the 
final state subject to the constraints of isospin addition and
charge conservation, and
the strengths of which ought to be computable in QCD in principle.
Indeed the $\theta_i$ of Kaufmann and Gibbs is precisely this
set for $\pi N$ scattering.\footnote{In the following,
we shall refer to the framework in which we determine the
relevant set of operators for pion photoproduction,
as the analogue of the framework of Kaufmann and
Gibbs for $\pi N$ scattering, in recognition of their pioneering
approach.}  

In the effective low-energy theory, it should also be possible to
recast all the operators arising in chiral perturbation theory
into combinations of the $\theta_i$ of Kaufmann and Gibbs with
calculable coefficients in terms of the low-energy constants and
quark masses, for partial results, see ref.~\cite{BAJLab}.

Bernstein~\cite{Bernstein3,Bernstein1,Bernstein2,Bernstein4} in a series of
publications has pointed out that pion photoproduction is an
ideal setting for probing isospin violation in the $\pi N$ sector,
in the reaction $\gamma p\to \pi N$.
[For recent experimental information on pion photoproduction, 
see, e.g.~\cite{A1COLLABORATION}.]  
It may be recalled that in the
isospin symmetric limit it is described in terms of 
two isospin amplitudes corresponding to $I=1/2,\, 3/2$.
A celebrated result associated with pion photoproduction
is that the pion photoproduction phases
are indeed the $\pi N$ phase shifts in this limit, which 
goes under the name of the Fermi-Watson theorem~\cite{FermiWatson}, when
the electromagnetic interaction is retained to leading order.
Bernstein has generalized the theorem in the presence of isospin
violation when there are three open channels, and on treating
quantities proportional to the isospin violating quantity
$m_d-m_u$ on par with quantities of $O(e)$, where $e$ is the
electronic charge.  The isospin violation due to the charged
and neutral pion mass difference, a quantity of $O(e^2)$ is
put in by hand, as is the elastic phase shift $\delta_\gamma$.

One of the important objectives of the present work is to 
provide an operator framework for the analysis of hadronic
isospin violation that feeds into pion photoproduction.
Stated differently, we provide a framework for pion photoproduction, 
analogous to that of Kaufmann and Gibbs for $\pi N$ scattering.
In practice this turns out to be straightforward, and 
requires an extension due to inclusion of neutron targets.  
There is considerable amount of data available for
this case as well, see, e.g.~\cite{Arndt1} and references therein.  

We note here that some authors have reported many results on 
the subject of (neutral) pion photoproduction
in the isospin conserving case~\cite{Ulfneutral}.  It is likely that isospin
violation is also worked out in chiral perturbation theory for
pion photoproduction, and it would be beneficial to cast the
results of those computations in terms of the operators presented
in this work.

In Sec.~\ref{FW} we review the proposal of Bernstein
by providing a setting for his version of the the Fermi-Watson theorem 
from a general approach to the unitarity conditions found in the 
literature~\cite{Oka,Henley}. 
This provides a unified framework for inequivalent representations
that Bernstein has considered in his treatment of the problem.
This would also benefit us, for we shall propose that a determination
of photoproduction multipole amplitudes from our operator approach
can then be fed back via the Fermi-Watson theorem to resolve
the discrepancies in the $\pi N$ phase shift analyses.

We will then proceed to describe the construction of the
operators that enter the photoproduction process and classify
the terms according to their tensorial properties in Sec.~\ref{OA}.
We will present expressions for the transition amplitudes
which will explicitly demonstrate the isospin violation.
We will propose that these ought to be the basis for the
analysis of the multipoles of pion photoproduction amplitudes.
We shall then compare the determination of isospin violation
in such an analysis, with that from the Fermi-Watson approach.
We shall finally point out that a phase shift analysis that
results from our approach could be fed back into the $\pi N$ system
via the Fermi-Watson theorem to resolve the discrepancy in those
analyses.

Of special interest is the possibility of carrying out
polarization measurements.  In particular, there is the
Jefferson Laboratory Letter of Intent~\cite{loi} (LOI) which proposes
to carry out photoproduction experiments at high precision
to determine better resonance parameters.  We 
provide a brief discussion on 
the significance of these measurements in
constraining isospin violation considered here in Sec.~\ref{POL}.

We recall that an analogous situation arose in the past 
when it was suggested that one may be able to observe an isotensor
contribution to the electromagnetic current.  This possibility is
not disallowed in the standard model at higher order, and was
considered seriously 
in~\cite{WeinbergTreiman,DombeyKabir}, while non-vanishing
evidence for its effect on photoproduction
was also reported in the past~\cite{SandaShaw}.
However, later experiments provided null results,
for a review, see, e.g. ref.~\cite{Kajikawa}.
We conclude by pointing out that the new polarization measurements
that are planned may be used to better constrain the isotensor
contributions.  

A summary is provided in Sec.~\ref{SUM}.
Finally in Appendix~\ref{AppA} we recall the main features of
the original formalism of Kaufmann and Gibbs for $\pi N$ scattering
as this process is so closely allied to pion photoproudction, and in
Appendix~\ref{AppB} we present the contributions of the amplitudes
including the possible isotensor contributions to the electromagnetic
currents to the reactions of interest and briefly describe
the special role that is played by the $\Delta(1232)$ resonance
in constraining the isotensor amplitude.

\section{Fermi-Watson theorem approach to isospin violation}\label{FW}
In this section we review the Fermi-Watson theorem approach to isospin 
violation.  Our treatment is based on the approach of Oka~\cite{Oka}
and that of Henley~\cite{Henley}.  We present expressions presented
by the latter, in a notation suited to our needs.  
For a three coupled channel $S-$matrix, in which we have
weak coupling between one channel denoted here by
$\gamma$, with the other two denoted by
$a,\, b$, where the channels are mutually absorptive, the
following representation holds for the partial waves of the
scattering:                                                           
\[ S^H=\left( \begin{array}{c c c}
\eta_{\gamma} e^{2 i \delta_\gamma} & 
                                      i \sqrt{\eta_a \eta_\gamma} S_{a\gamma}^H
                                        e^{i(\delta_a+\delta_\gamma)} &
                                      i \sqrt{\eta_b \eta_\gamma} S_{b\gamma}^H
                                        e^{i(\delta_b+\delta_\gamma)} \\
i \sqrt{\eta_a \eta_\gamma} S_{a\gamma}^H
                                        e^{i(\delta_a+\delta_\gamma)} &
\sqrt{\eta_a^2-\eta_a\eta_b \rho^2} 
e^{2 i \delta_a} & 
                                      i \sqrt{\eta_a \eta_b}\rho 
                                        e^{i(\delta_a+\delta_b)}  \\
                                      i \sqrt{\eta_b \eta_\gamma} S_{b\gamma}^H
                                        e^{i(\delta_b+\delta_\gamma)} &
i \sqrt{\eta_a \eta_b} \rho 
                                        e^{i(\delta_a+\delta_b)} &
\sqrt{\eta_b^2-\eta_a\eta_b \rho^2} 
 e^{2 i \delta_b} \\
\end{array} \right), \]
provided:
\begin{equation}
S_{a\gamma}^H=\epsilon_{a\gamma}^H+i \epsilon_{b\gamma}^H{\eta_a\rho
		\over \eta_\gamma+\sqrt{\eta_b^2-\eta_a\eta_b \rho^2}},
\end{equation}
and
\begin{equation}
S_{b\gamma}^H=\epsilon_{b\gamma}^H+i \epsilon_{a\gamma}^H{\eta_b\rho
		\over \eta_\gamma+\sqrt{\eta_a^2-\eta_a\eta_b \rho^2}}.
\end{equation}
In the above, $\epsilon_{a\gamma}^H,\, \epsilon_{b\gamma}^H$ represent
the matrix elements for the transitions, $\rho$ is the absorption
parameter in the $2\times 2$ subsector spanned by $a,\, b$.   

Bernstein has presented expressions for two cases, in the
limit when $\eta_i,i=\gamma,a,b$ are all equal to unity.
These correspond to the cases when 

\medskip

(A) $a=0,\, b=c$, for the case
of elastic and charge exchange scattering in which
the three channels of interest are
$\gamma p,\, \pi^0 p,\, \pi^+ n$~\cite{Bernstein1}, and 

\medskip

(B) $a=1,\, b=3$
which represent the the value $2 I$, where $I$ is the definite
isospin in the $\pi N$ system, where the
three channels of interest are 
$\gamma N, (\pi N)^{2I=1},$ $ (\pi N)^{2I=3}$~\cite{Bernstein2}.  

\medskip

In case (A), and in the limit of unity elasticities, $\rho$ is
identified with $\sin\phi$ in ref.~\cite{Bernstein1} and the
transition matrix elements are the corresponding multipole amplitudes
for pion photoproduction.  
For completeness we reproduce the S-matrix given therein,
for the three channels 
$\gamma p,\, \pi^0 p,\, \pi^+ n$:
\[ \left( \begin{array}{c c c}
e^{2 i \delta_\gamma} & i M'_0 & i M'_c \\ 
i M'_0 &  \cos\phi e^{2 i \delta_0} &  i \sin\phi e^{i(\delta_0+\delta_c)}  \\
i M'_c & i \sin\phi e^{i(\delta_0+\delta_c)} & \cos\phi e^{2 i \delta_c}
\end{array} \right) \] 
In this limit the multipoles for pion photoproduction read:
\begin{eqnarray*}
M'_0=e^{i(\delta_\gamma+\delta_c)}\left[A'_0 \cos(\phi/2) + i A'_c
                                   \sin(\phi/2) \right] \\
M'_c=e^{i(\delta_\gamma+\delta_c)}\left[A'_c \cos(\phi/2) + i A'_0
                                   \sin(\phi/2) \right]. 
\end{eqnarray*}
In the above $A'_0,\, A'_c$ are quantities proportional to the
multipole matrix elements for the charge non-exchange and charge
exchange scattering respectively.
Bernstein proceeds to relate the quantities above
to multipole amplitudes of pion photoproduction.
In the near threshold region, we have $\cos({\phi/2})\to 1$ and
here it is now possible define a quantity
\begin{eqnarray*}
& \displaystyle 
\beta \simeq
E_{0+}(\gamma p \to \pi^+ n) a_{cex}(\pi^+ n \to \pi^0 p), & 
\end{eqnarray*} 
where $E_{0+}$ is the multipole moment and $a_{cex}$ is a
$\pi N$ scattering length.
This quantity is used to demonstrate the unitarity cusp associated
with the two-step process $\gamma p \to \pi^+ n \to \pi^0 p$, in the
limit of isospin conservation, barring the pion mass difference.
Furthermore, in this limit Bernstein points out that 
the presence of isospin violation denoted by $\delta a_{cex}$ may be
detected.  

It must be pointed out that away from the threshold region, 
the limit $\cos(\phi/2) \to 1$ no longer holds.
Our operator construction
of the next section may be used to demonstrate isospin violation
away from threshold as well.

\medskip

In case (B), the result is presented for the case where
$\rho=\sin\psi$, where $\psi$ is a small quantity.  
This corresponds to the S-matrix for the channels
$\gamma N, (\pi N)^{2I=1}, (\pi N)^{2I=3}$:  
\[ \left( \begin{array}{c c c}
e^{2 i \delta_\gamma} & i M_1 & i M_3 \\
i M_1 & \cos\psi e^{2 i \delta_1} & i \sin\psi e^{i(\delta_1+\delta_3)} \\
i M_2 & i \sin\psi e^{i(\delta_1+\delta_3)} & \cos\psi e^{2 i \delta_3} 
\end{array} \right) \] 
The unitarity condition then yields:
\begin{eqnarray*}
M_1=e^{i(\delta_\gamma+\delta_1)}\left[A_1 \cos(\psi/2) + i A_3
                                   \sin(\psi/2) \right] \\
M_3=e^{i(\delta_\gamma+\delta_3)}\left[A_3 \cos(\psi/2) + i A_1
                                   \sin(\psi/2) \right]. 
\end{eqnarray*}
In the above $A_1,\, A_3$ are quantities proportional to the
multipole matrix elements for the amplitudes of definite isospin
in the absence of final state interactions and isospin violation. 
Bernstein also sets for this case, $\delta_\gamma=0$.  
Using experimental information based on two independent
analyses of $\pi N$ scattering, Bernstein concludes
that at a pion kinetic energy of about 40 MeV,
$\psi\simeq 0.010\pm 0.004$~\cite{Bernstein2}.
In contrast, it is hoped that the operator approach which is to be
described in the next section can assist in unambiguously
demonstrating isospin violation, without taking recourse
to any information from the $\pi N$ sector.
Furthermore, this coupled channel analysis is valid only
below the $2\pi$ threshold.

\vfill

\section{Operator approach to isospin violation}\label{OA}
The traditional analysis of pion photoproduction,
see ref.~\cite{DT}, relies on two 
assumptions: 

\noindent (a) that the electromagnetic current transforms as
\begin{eqnarray*}
& \displaystyle {1\over 2} \left(f^s + f^v \tau_0\right), &
\end{eqnarray*}
{\it viz.} an isoscalar and an isovector part, 
and, and that;

\noindent (b) there is no isospin violation in the hadronic
system, due to which the interaction in isospin is
proportional to an operator that transforms  an isoscalar:
\begin{equation}\label{scalar}
O_S\equiv{\bf \tau}\cdot {\bf \Phi}.
\end{equation}
In the above, ${\bf \tau}\equiv(\tau_0,\tau_1,\tau_2)$ is an isovector
containing the Pauli matrices and ${\bf \Phi}$ is an isovector
containing the pions.

In the past, assumption (a) has been questioned (and
it has been shown that even in the standard model, 
at higher order in the electromagnetic field, this
assumption is violated).  This is the basis of the
isotensor contribution to the electromagnetic current,
see Appendix~\ref{AppB},\footnote{
The determination of this contribution to the amplitude is a tremendous 
experimental challenge.  
We shall
discuss this further in Sec.~\ref{POL}.  In Appendix B, we also
provide a short discussion on the mechanism for the
relevant isotensor contributions.}   
which transforms as
\begin{eqnarray*}
& \displaystyle {f^t\over 2\sqrt{15}} \left(\tau_1 +\tau_2 -2 \tau_0\right). &
\end{eqnarray*}

There has been no treatment of a departure from assumption (b)
in general in the literature.  In fact, by providing all possible
isospin violating terms in this context, here we are providing
the general operator framework accounting for strong
isospin violation in the process. This amounts to providing 
the counterpart for pion photoproduction, of the framework of
Kaufmann and Gibbs for $\pi N$ scattering.

Isospin violation in the hadronic system can arise from
the most general term of the type $\tau_i \Phi_j, \, i,j=0,1,2$.
The nine possible combinations can be organized into a
scalar $O_S$, a vector whose components are given by
\begin{eqnarray*} 
-i \epsilon^{i j k} {\tau}_j {\Phi}_k,
\end{eqnarray*}
and a traceless symmetric tensor whose components are
\begin{eqnarray*}
& \displaystyle 
({ \tau}_i { \Phi}_j +{ \tau}_j { \Phi}_i)(1-\delta^{ij}), \, 
{ \tau}_1 { \Phi}_1 -{ \tau}_2 { \Phi}_2, \, 
{ \tau}_1 { \Phi}_1 +{ \tau}_2 { \Phi}_2  -2 
                           { \tau}_0 { \Phi}_0. \, &  
\end{eqnarray*}
Of the operators listed above, the $i=0$ component of the vector operator
alone, and the last of the tensor components listed above alone conserve 
electric charge.  Therefore we can introduce 2 operators:  
\begin{eqnarray}
& \displaystyle O_V\equiv -i (\tau_1 \Phi_2 - \tau_2 \Phi_1) & \\
& \displaystyle O_T \equiv 
{ \tau}_1 { \Phi}_1 +{ \tau}_2 { \Phi}_2  -2 
                           { \tau}_0 { \Phi}_0. \, &  
\end{eqnarray}
The set $O_S,\, O_V,\, O_T$ for pion photoproduction, is the counterpart
of the set $\theta_i,\, i=1,...10$ of $\pi N$ scattering in the
Kaufmann and Gibbs framework (see Appendix~\ref{AppA}). 
It may be reiterated that $O_S$ is isospin conserving while the
other two, $O_V,\, O_T$ are isospin violating.

We begin by recalling that the overall matrix element for the scalar case  
involves the amplitudes that we shall denote by
 $A^{(-)}_S$, $A^{(+)}_S$ and $A^{(0)}_S$ 
when the Pauli matrices appearing in the interaction of the
nucleon with the photon and pion are arranged as
\begin{equation}
\left({1\over 2} A^{(-)}_S [\tau_i, \tau_0] + {1\over 2}
A^{(+)}_S \{\tau_i,\tau_0\}+ A^{(0)}_S\right) \Phi_i.
\end{equation}
In analogy therefore, the new operators contribute to the matrix
element for the cases of vector and tensor through amplitudes
denoted by $A^{(-)}_R,\, A^{(+)}_R,\, R=V,T$ 
associated with the commutator, anti-commutator accompanying
$f^v$ and $A^{(0)}_R,\, R=V,T$ accompanying $f^s$. 

The contributions of these amplitudes to the physical reactions may
now be evaluated in a straightforward manner which then reads
for the reactions denoted by $R_a,\, a=1,2,3,4$, and defined
below.  They read: 
\begin{small}
\begin{eqnarray}
 & \displaystyle
    R_1:  \, T(\gamma n\to \pi^0 n) = & \nonumber \\ 
 & \displaystyle -A^{(0)}_S +  A^{(+)}_S
                                         + 2 A^{(0)}_T - 2 A^{(+)}_T, & \\
& \displaystyle R_2: \, T(\gamma p\to \pi^0 p) = &  \nonumber \\ 
 & \displaystyle A^{(0)}_S +  A^{(+)}_S
                                         - 2 A^{(0)}_T - 2 A^{(+)}_T, & \\
& \displaystyle R_3: \,  T(\gamma n\to \pi^- p) = & \nonumber \\
 & \displaystyle
   \sqrt{2} A^{(0)}_S -\sqrt{2} A^{(-)}_S - \sqrt{2} A^{(0)}_V +
    \sqrt{2} A^{(-)}_V +\sqrt{2} A^{(0)}_T - \sqrt{2} A^{(-)}_T, & \\
& \displaystyle R_4: \, T(\gamma p\to \pi^+ n) = & \nonumber \\
 & \displaystyle
   \sqrt{2} A^{(0)}_S +\sqrt{2} A^{(-)}_S + \sqrt{2} A^{(0)}_V +
    \sqrt{2} A^{(-)}_V + \sqrt{2} A^{(0)}_T + \sqrt{2} A^{(-)}_T. & 
\end{eqnarray}
\end{small}
We take this opportunity to suggest that this set of amplitudes be
the basis for the analysis of pion photoproduction multipole analysis.
In this manner, isospin violation in the hadronic sector could be
probed with no recourse to the Fermi-Watson theorem.
The amplitudes $A_R^{(\pm)},
\, A_R^{(0)},\, R=V,T$ get contributions due to $(m_d-m_u)\neq 0$. 
The vector amplitudes receive contributions at leading order in this quantity,
while the tensor will receive contributions only at higher order.

We may infer from this that the analogue of the triangle relation of
Kaufmann and Gibbs (see Appendix~\ref{AppA}) for pion photoproduction
reads:
\begin{eqnarray}\label{newrelation}
& \displaystyle T(\gamma n\to \pi^- p) + T(\gamma p \to \pi^+ n)
=-\sqrt{2}\left(T(\gamma n\to \pi^0 n) - T(\gamma p \to \pi^0 p)\right) &
\end{eqnarray}
in the absence of isospin violation, {\it viz,} when all the
amplitudes due to $O_V,\, O_T$ are set to zero.

In light of the expressions above,
it may be seen that indeed one cannot probe the
vector like isospin violating interactions without a charge
exchange reaction involving the nucleons.  This is in accordance
with the observations of Weinberg~\cite{Weinberg1} and those of
Bernstein~\cite{Bernstein3,Bernstein1,Bernstein2}.  
On the other hand, it is possible
to observe isospin violating interactions of the tensor type in
neutral pion production.  Such an interaction is not likely to be
important in the low-energy regime where the isospin violating
contribution can be coupled
at leading order only in a vectorlike manner.  However, our
framework opens up the possibility of probing isospin violation
at higher energies by considering the reactions above.

\section{Polarization experiments}\label{POL}
Bernstein~\cite{Bernstein3,Bernstein2,Bernstein1} has pointed
out repeatedly the availability of polarized targets/beams
would significantly enhance the capacity of experiments to
probe isospin violation.  The main reason for this is that
polarization affords the possibility of measuring the multipole amplitude
${\rm Im}(E_{0+})$ (for notation see ref.~\cite{DT}) in the
near threshold region.  
In this regard, we draw attention to the Jefferson Laboratory 
LOI~\cite{loi} where it has been proposed to carry out photo production
experiments using target/beam polarization.  It is expected
that there will be high statistics experiments, including also
neutron targets (the proposal here involves both deuteron as
well as carbon targets).  
Indeed, the measurements of the low multipoles of photoproduction
amplitudes could be used to study the deviations from the
isospin conserving relation given in eq.~(\ref{newrelation}).
It should also be possible to determine, process by process,
the contributions of the isospin violating amplitudes to
the phases of the multipole amplitudes.  Indeed, there is already
data from the experiment E94-104~\cite{E94104} at high energies,
which could possibly be analyzed for isospin violating effects
at these energies.

Another of the important objectives set out in the
Jefferson Laboratory LOI~\cite{loi} is to determine
better the parameters of the resonance denoted by $P_{33}(1232)$.
In this regard,
we now turn to the issue of the determination of the isotensor
contribution to the electromagnetic current.  (Note also that our
amplitude $A_T^{(0)}$, upto a numerical factor is not distinguishable
from the isotensor contribution
to the electromagnetic current.)   The determination of this
amplitude is a very challenging one from an experimental 
point of view due to the contributions from $A_S^{(0)}$
in the non-resonant region.  However,
one place where a clear signature can be seen
is at an energy corresponding to the production of an
$I=3/2$ resonance, the $\Delta(1232)$, where the isoscalar part of the
current makes no contribution, and the production amplitude
would involve only the isovector and isotensor parts, resulting
in an interference between the two contributions.  Here,
Sanda and Shaw~\cite{SandaShaw} find evidence for an isotensor
contribution to the electromagnetic current from data obtained
with polarized photons.  However, from analysis of later experimental data
there have been  null
results presented in the literature.
In ref.~\cite{PRLTA.33.1500} an experiment with tagged photons
found no evidence for isotensor component to the crosssection, while
ref.~\cite{Fujii} reports results from an 
experiment that observes the differential crosssection
at several different angles, which also found no evidence. 
Null results are also reported in ref.~\cite{Kajikawaref16},
based on experiments with polarized photons.  Here it has
been pointed out that target asymmetries could play a role
in resolving ambiguities.  
In the light of the proposals presented in the LOI, and due
to the likelihood of the availability of polarization and other facilities
at Jefferson Laboratory
detailed therein, the upcoming experiments there
could play a crucial role in settling this question.\footnote{
In this regard,
it should be noted that pion electroproduction
experiments have been consistently seeing evidence for an isotensor
amplitude.  Data from the recent SLAC experiments NE11 and E133
yield~\cite{Stuart}, for the ratio 
${\sigma_n/ \sigma_p}$
for the crosssections on neutron and proton targets, $0.72\pm
0.09$, to be contrasted with the value from older data of
K\"obberling~\cite{Kobberling}, of $0.91\pm 0.03$, at the $\Delta(1232)$ 
resonance.  This ratio should be unity in the absence of isotensor
contributions.}

\section{Discussion and Summary}\label{SUM}
In this work, we have revisited the issue of observing
isospin violation in the hadronic sector from pion photoproduction.
We have pointed out that the application of the Fermi-Watson
theorem is one approach that has been considered in the past.
We have presented a comparison of different techniques used
to arrive at the pertinent expressions by appealing to general
treatments.  We have pointed out that there is a general operator
approach also to the phenomenon, which we have worked out here.
In essence, it is the counterpart of the Kauffman and Gibbs construction
for $\pi N$ scattering.  

The operator approach described here, yields a set of amplitudes
for pion photoproduction which should be the basis for the determination
of the multipole amplitudes for the processes of interest.  The
resulting phase shifts can then be inserted into a Fermi-Watson like
system to provide a set of constraints for $\pi N$ phase shift analyses
which are in mutual disagreement at the moment.  The Fermi-Watson approach
of Bernstein uses well known $\pi N$ phase shift analyses to establish
isospin violation in photoproduction.  This latter is also constrained
to be valid only below the $2 \pi$ threshold, and is to leading order
in the electric charge.  The treatment presented in Case (A) of Bernstein
requires the $\pi N$ scattering length as an input to demonstrate
isospin violation at the photoproduction threshold, while the treatment
in Case (B) requires $\pi N$ phase shift analyses as an input.  
Our treatment does not require these inputs.

We have considered the virtues of 
polarization experiments and have pointed out that at the
upcoming facilities, one may constrain isotensor contributions
to the electromagnetic current, expected to arise at higher orders
better.  Our work is likely to be a useful platform for the construction
of results from effective Lagrangians and a basis for analysis of
crosssections which can be used to constrain isospin violation in
the hadronic sector.  Finally, we point out here that a determination
of photoproduction amplitudes including the effects due to
the isospin violating operators presented here, could then be used
via the Fermi-Watson theorem to resolve the discrepancies in
the $\pi N$ phase shift analyses.

\section{Acknowledgements}  
We thank A.~M.~Bernstein for providing his contribution to
the AIP Conference Proceedings of ref.~\cite{Bernstein4}, 
P. B\"uttiker, C.~R.~Das, J.~Goity for discussions,
W.~B.~Kaufmann for correspondence (including on the matter of the
entries of Table IV of ref.~\cite{KG}, see footnote [4]), 
H.~Leutwyler for discussions and comments on a preliminary
version of the manuscript, U-G.~Mei{\ss}ner, B.~Moussallam
and D.~Sen for discussions.  We thank K. Shivaraj for a careful
reading of the manuscript.
The hospitality of the Theory Group,
Thomas Jefferson National Accelerator Facility,
USA at the time this work was initiated is acknowledged.  
The work is supported in 
part by the CSIR under scheme number 03(0994)/04/EMR-II 
and by the Department of Science and Technology.


\setcounter{section}{0}
\renewcommand{\thesection}{\Alph{section}}
\section{Formalism of Kaufmann and Gibbs for $\pi N$
scattering}\label{AppA}
For the ten reactions (listed
in Table I of ref.~\cite{KG}) of interest, 
a general analysis of isospin violation
in terms of a set of 10 standard operators designated
$\theta_i,\, i=1,...,10$ (listed in Table II of
ref.~\cite{KG}) is presented.
The matrix elements for
all these operators are listed (see Table III of ref.~\cite{KG})  
and the result is also presented in the isospin basis
(see Table IV of ref.~\cite{KG}\footnote{We point out here that 
all the signs 
corresponding to the entries of $\theta_9$ and $\theta_{10}$
in Table IV of ref.~\cite{KG} need to be consistently reversed.}). 
We note here some of the features of the work: 
\begin{enumerate}
\item{$\theta_{1,5}$ are isospin conserving, $\theta_{4,6,(10)}$
violate isospin but are invariant under charge reflection,
$\theta_{3,7,(9)}$ conserve neither isospin nor charge reflection,
$\theta_8$ besides not conserving isospin and charge reflection, only
connects $I=3/2$ states}.
\item{In the elementary examples of isospin violation, the
combination
\begin{equation}
\theta_3 -\sqrt{{1\over3}} \theta_5 + \sqrt{{2\over3}} \theta_6
\end{equation}
represents the Coulomb interaction and that this combination gives
the product of the nucleon and pion charge.}
\item{
$\pi^0-\eta$ mixing, a quantity that receives contributions at
leading order in $(m_d-m_u)$ transforms as
\begin{equation}\label{pi0eta}
\sqrt{{8\over 9}} \theta_2 + \sqrt{{40\over 9}} \theta_7.
\end{equation}
}
\item{
The triangle identity, which holds in the
isospin conserving limit, can be expressed as
\begin{equation}
T (\pi^+ p \to \pi^+ p) -T (\pi^- p \to \pi^- p) = \sqrt{2}\, 
T (\pi^+ p \to \pi^0 n).
\end{equation}
}
\end{enumerate}
In the work of Kaufmann and Gibbs a treatment of a final state theorem
is presented, which allows one to transform certain $I=1$ operators into
other $I=1$ operators by left and right multiplication by isospin
conserving operators.  In particular, it could turn an operator that
transforms with the transformation law of $\rho-\omega$ mixing
($\theta_3$) into
one that has the transformation law of $\pi^0-\eta$ mixing. 
\bigskip

\renewcommand{\thesection}{\Alph{section}}
\section{Isotensor contributions}\label{AppB}
We begin by recalling that 
in the limit of isospin conservation in the hadron sector, 
we have the isospin relations for the amplitudes
$t^{2I}, \, I=3/2, 1/2$ reading
\begin{eqnarray*}
A^3= A^{(+)}_S - A^{(-)}_S  \\
A^1= A^{(+)}_S +2 A^{(-)}_S 
\end{eqnarray*}
Note that $A^0\equiv A^{(0)}_S$ contributes to $I=1/2$ amplitude.
The admission of an isotensor operator 
can lead up to $I=3/2$ state, when
represented by an amplitude $A^2$.
These together yield for the processes
of interest~\cite{SandaShaw}:
\begin{eqnarray*}
& 
R_1: -A^0+A^1/3 + 2 A^2/\sqrt{15}  +2 A^3/3, & \\
& 
R_2:  A^0+A^1/3 - 2 A^2/\sqrt{15}  +2 A^3/3, & \\
& 
R_3: \sqrt{2}\left( A^{ 0 } - A^1/3 + A^2/\sqrt{15}  +
		A^3/3\right), & \\
& 
R_4: \sqrt{2} \left( A^{ 0 } + A^1/3 +A^2/\sqrt{15}  -
                            A^3/3\right). & 
\end{eqnarray*}
It may also be noted that the amplitude $A^2$ contributes
to the $R_i, \, i=1,2,3,4$ in the same way as
$\sqrt{15} A^{(0)}_T$.  
These amplitudes are the basis of the analysis of
photoproduction amplitudes in ref.~\cite{SandaShaw}.
In this work, the multipole amplitudes $M_{1+}^i,\, i=0,1,2,3$ have been studied
in detail.

We present here some salient features of the possibility of
detecting the signature of the isotensor amplitude from the
formation of the resonance 
$\Delta(1232)$~\cite{DombeyKabir,SandaShaw} (see also
refs.~\cite{GittelmanSchmidt,DonnachieShaw}).
At the resonance, the isoscalar amplitude does not contribute
to the crosssection except for the nonresonant background.
If, for example, the dominant multipoles $M_{1+}$ are being probed,
then the presence of the isotensor would lead to an
interference term proportional to ${\rm Re} (M_{1+}^2 M_{1+}^3)$.
The model for the isotensor term which is the basis of
the analysis of Sanda and Shaw~\cite{SandaShaw} is written down in 
the static model of Chew et al.~\cite{CGLN}, by introducing 
an isotensor $\gamma\Delta n$ coupling, and required 
the resulting multipole moment to verify a fixed-$t$ dispersion
relation.  This allows for the isotensor interaction to participate
in the resonance production.


\vfill

\begin{thebibliography}{abc}
\bibitem{Key}
  S.~Capstick {\it et al.},
  arXiv:hep-ph/0012238.

\bibitem{MNS}
  G.~A.~Miller, B.~M.~K.~Nefkens and I.~Slaus,
  Phys.\ Rept.\  {\bf 194}, 1 (1990).

\bibitem{Gardestig2}
  A.~Gardestig,
  AIP Conf.\ Proc.\  {\bf 768}, 45 (2005)
  [arXiv:nucl-th/0409025].

\bibitem{PRLTA.91.142302}
  E.~J.~Stephenson {\it et al.},
  Phys.\ Rev.\ Lett.\  {\bf 91}, 142302 (2003)
  [arXiv:nucl-ex/0305032].

\bibitem{PRLTA.91.212302}
  A.~K.~Opper {\it et al.},
  Phys.\ Rev.\ Lett.\  {\bf 91}, 212302 (2003)
  [arXiv:nucl-ex/0306027].

\bibitem{Gardestig}
  A.~Gardestig {\it et al.},
  Phys.\ Rev.\ C {\bf 69}, 044606 (2004)
  [arXiv:nucl-th/0402021].

\bibitem{Nuclear1}
  U.~van Kolck, J.~L.~Friar and T.~Goldman,
  Phys.\ Lett.\ B {\bf 371}, 169 (1996)
  [arXiv:nucl-th/9601009].

  U.~van Kolck,
  Nucl.\ Phys.\ A {\bf 631}, 56C (1998)
  [arXiv:hep-ph/9707228].

  U.~van Kolck,
  arXiv:hep-ph/9711222.


  E.~Epelbaum {\it et al.}, 
  AIP Conf.\ Proc.\  {\bf 603}, 17 (2001)
  [arXiv:nucl-th/0109065].

\bibitem{Nuclear2}
  S.~A.~Coon,
  arXiv:nucl-th/9903033.

  M.~Walzl, U.-G.~Mei{\ss}ner and E.~Epelbaum,
  Nucl.\ Phys.\ A {\bf 693}, 663 (2001)
  [arXiv:nucl-th/0010019].

\bibitem{Ericson}
  T.~E.~O.~Ericson,
  arXiv:hep-ph/0504258.

\bibitem{Gibbs_Li_Kaufmann}
  W.~R.~Gibbs, L.~Ai and W.~B.~Kaufmann,
  Phys.\ Rev.\ Lett.\  {\bf 74}, 3740 (1995).

  W.~R.~Gibbs, L.~Ai and W.~B.~Kaufmann,
  Phys.\ Rev.\ C {\bf 57}, 784 (1998)
  [arXiv:nucl-th/9704058].

\bibitem{Matsinos}
  E.~Matsinos,
  Phys.\ Rev.\ C {\bf 56}, 3014 (1997).

\bibitem{Weinberg1}
  S.~Weinberg,
  Trans.\ New York Acad.\ Sci.\  {\bf 38}, 185 (1977).

\bibitem{GL_report}
J.~Gasser and H.~Leutwyler,
  Phys.\ Rept.\  {\bf 87}, 77 (1982).

\bibitem{CHPTReviews}  
 H.~Leutwyler,
  arXiv:hep-ph/9406283.

  J.~Bijnens, G.~Ecker and J.~Gasser,
  arXiv:hep-ph/9411232.

  S.~Scherer,
  arXiv:hep-ph/0210398.

\bibitem{Meissneretal}
  U.-G. ~Mei{\ss}ner and S.~Steininger,
  Phys.\ Lett.\ B {\bf 419}, 403 (1998)
  [arXiv:hep-ph/9709453].

  N.~Fettes, U.-G.~Mei{\ss}ner and S.~Steininger,
  Phys.\ Lett.\ B {\bf 451}, 233 (1999)
  [arXiv:hep-ph/9811366].

  N.~Fettes and U.-G.~Mei{\ss}ner,
  Phys.\ Rev.\ C {\bf 63}, 045201 (2001)
  [arXiv:hep-ph/0008181].

\bibitem{KG}
  W.~B.~Kaufmann and W.~R.~Gibbs,
  Annals Phys.\  {\bf 214}, 84 (1992).

\bibitem{BAJLab}
B. Ananthanarayan, Jefferson Laboratory Technical note, JLAB-TN-03-33 

\bibitem{Bernstein3}
  A.~M.~Bernstein,
  PiN Newslett.\  {\bf 11}, 66 (1995).

  A.~M.~Bernstein,
  PiN Newslett.\  {\bf 13}, 37 (1997).

\bibitem{Bernstein1}
  A.~M.~Bernstein,
  Phys.\ Lett.\ B {\bf 442}, 20 (1998)
  [arXiv:hep-ph/9810376].

\bibitem{Bernstein2}
  A.~M.~Bernstein,
  PiN Newslett.\  {\bf 15}, 140 (1999).

\bibitem{Bernstein4}
  A.~M.~Bernstein,
{\it 
Electromagnetic pion production: From Yukawa to Goldstone}, 
 AIP Conference Proceedings -- June 16, 2000 -- Volume 520, Issue 1, pp. 254-270


\bibitem{A1COLLABORATION}
  H.~Merkel {\it et al.},
  Phys.\ Rev.\ Lett.\  {\bf 88}, 012301 (2002)
  [arXiv:nucl-ex/0108020].

\bibitem{FermiWatson}
  E.~Fermi,
{\it Prepared for 4th Annual Rochester Conference on High-Energy and Nuclear Physics, Rochester, New York, 25-27 Jan 1954}

  K.~M.~Watson,
  Phys.\ Rev.\  {\bf 95}, 228 (1954).

\bibitem{Arndt1}
  R.~A.~Arndt, W.~J.~Briscoe, I.~I.~Strakovsky and R.~L.~Workman,
  Phys.\ Rev.\ C {\bf 66}, 055213 (2002)
  [arXiv:nucl-th/0205067].

\bibitem{Ulfneutral}
  V.~Bernard, N.~Kaiser and U.-G.~Mei{\ss}ner,
  Eur.\ Phys.\ J.\ A {\bf 11}, 209 (2001)
  [arXiv:hep-ph/0102066].

  H.~Krebs, V.~Bernard and U.-G.~Mei{\ss}ner,
  Nucl.\ Phys.\ A {\bf 713}, 405 (2003)
  [arXiv:nucl-th/0207072].

  H.~Krebs, V.~Bernard and U.-G.~Mei{\ss}ner,
  Eur.\ Phys.\ J.\ A {\bf 22}, 503 (2004)
  [arXiv:nucl-th/0405006].


\bibitem{Oka}
  T.~Oka,
  Prog.\ Theor.\ Phys.\  {\bf 66}, 977 (1981).

\bibitem{Henley}
  E.~M.~Henley,
  Nucl.\ Phys.\ A {\bf 483}, 596 (1988).

\bibitem{loi}
R.~Arndt et al., {\it Pion Photoproduction from a Polarized Target},
Letter of Intent to Jefferson Lab PAC 22, 2001.

\bibitem{WeinbergTreiman}
  S.~Weinberg and S.~Treiman, Phys.\ Rev. {\bf 116}, 465 (1959).

\bibitem{DombeyKabir}
   V.~G.~Grishin {\it et al.},
   Yadern.\ Fiz.\ {\bf 4}, 126 (1966).

  N.~Dombey and P.~K.~Kabir,
  Phys.\ Rev.\ Lett.\  {\bf 17}, 730 (1966).


\bibitem{SandaShaw}
  A.~I.~Sanda and G.~Shaw,
  Phys.\ Rev.\ Lett.\  {\bf 24}, 1310 (1970).

  A.~I.~Sanda and G.~Shaw,
  Phys.\ Rev.\ D {\bf 3}, 243 (1971).

\bibitem{Kajikawa}
  R.~Kajikawa,
DPNU-8-76
{\it Review talk at INS Symposium on Electron and Photon Interactions in Resonance Region and on Related Topics, Tokyo, Japan, Nov 25-27, 1975}


\bibitem{DT}
  D.~Drechsel and L.~Tiator,
  J.\ Phys.\ G {\bf 18}, 449 (1992).

\bibitem{E94104}
  L.~Y.~Zhu {\it et al.}  [Jefferson Lab Hall A Collaboration],
  Phys.\ Rev.\ C {\bf 71}, 044603 (2005)
  [arXiv:nucl-ex/0409018].

\bibitem{PRLTA.33.1500}
  R.~W.~Clifft {\it et al.},
  Phys.\ Rev.\ Lett.\  {\bf 33}, 1500 (1974).

\bibitem{Fujii}
  T.~Fujii {\it et al.},
  Nucl.\ Phys.\ B {\bf 120}, 395 (1977).

\bibitem{Kajikawaref16}
  S.~Suzuki, S.~Kurakawa and K.~Kondo,
  Nucl.\ Phys.\ B {\bf 68}, 413 (1974).


\bibitem{Stuart}
  L.~M.~Stuart {\it et al.},
  Phys.\ Rev.\ D {\bf 58}, 032003 (1998)
  [arXiv:hep-ph/9612416].

\bibitem{Kobberling}
  M.~K\"obberling {\it et al.} 
  Nucl.\ Phys.\ B {\bf 82}, 201 (1974).


\bibitem{GittelmanSchmidt}
G.~Shaw,
  Nucl.\ Phys.\ B {\bf 3}, 338 (1967) 

  B.~J.~Gittelman and W.~Schmidt,
  Phys.\ Rev.\  {\bf 175}, 1998 (1968).

\bibitem{DonnachieShaw}
  A.~Donnachie and G.~Shaw,
  Phys.\ Rev.\ D {\bf 5}, 1117 (1972).

\bibitem{CGLN}
G.~F.~Chew, M.~L.~Goldberger, F.~E.~Low and Y.~Nambu,
  Phys.\ Rev.\  {\bf 106}, 1345 (1957).

\end{thebibliography}
\end{document}